\documentclass[runningheads]{llncs}
\usepackage[T1]{fontenc}

\usepackage{graphicx}
\usepackage{subcaption} %
\usepackage{adjustbox} %
\usepackage{booktabs}
\usepackage{hyperref}

\usepackage{xcolor}
\usepackage{tcolorbox}
\usepackage{tikz}
\usepackage{cleveref}
\usepackage{enumitem}

\definecolor{grey}{RGB}{89, 89, 89}
\definecolor{blue}{RGB}{38, 105, 202} %
\definecolor{red}{RGB}{204, 102, 119} %
\definecolor{turquoise}{RGB}{68, 170, 153} %

\newcommand{\greyCircled}[1]{\tikz[baseline=(char.base)]{
            \node[shape=circle,draw=grey,text=grey,inner sep=0.5pt,scale=0.85] (char) {#1};}}

\newcommand{\blueCircled}[1]{\tikz[baseline=(char.base)]{
            \node[shape=circle,draw=blue,text=blue,inner sep=0.5pt,scale=0.85] (char) {#1};}}

\tcbset{
  myboxstyle/.style={
    colback=turquoise!5,
    colframe=turquoise,
    colbacktitle=turquoise,
    boxrule=0.4pt,
    arc=1mm,
    left=6pt,
    right=6pt,
    top=6pt,
    bottom=6pt,
    fonttitle=\bfseries,
    width=\columnwidth,
  }
}

\newcommand{\todo}[1]{}

\newcommand{\change}[2]{#2}

\usepackage[normalem]{ulem}

\begin{document}
\title{Can Large Language Models Help Students Prove Software Correctness? An Experimental Study with Dafny}
\titlerunning{Can Large Language Models Help Students Prove Software Correctness?}

\author{Carolina Carreira\inst{1}%
\and
Álvaro Silva\inst{2}%
\and
Alexandre Abreu\inst{2}%
\and
Alexandra Mendes\inst{2}%
}
\authorrunning{C. Carreira et al.}

\institute{Carnegie Mellon University, USA, 
INESC-ID \& IST, University of Lisbon, Portugal \and INESC TEC, Faculty of Engineering, University of Porto, Portugal
}

\maketitle              %

\begin{abstract}
Students in computing education increasingly use large language models (LLMs) such as ChatGPT. Yet, the role of LLMs in supporting cognitively demanding tasks, like deductive program verification, remains poorly understood. This paper investigates how students interact with an LLM when solving formal verification exercises in Dafny, a language that supports functional correctness by allowing programmers to write formal specifications and automatically verifying that the implementation satisfies the specification. We conducted a mixed-methods study with master's students enrolled in a formal methods course. Each participant completed two verification problems, one with access to a custom ChatGPT interface that logged all interactions and the other without. 
We identified strategies used by successful students and assessed the level of trust students place in LLMs. 
Our findings show that students perform significantly better when using ChatGPT; however, performance gains are tied to prompt quality. 
We conclude with practical recommendations for integrating LLMs into formal methods courses more effectively, including designing LLM-aware challenges that promote learning.


\end{abstract}

\section{Introduction}

Large language models (LLMs), such as ChatGPT, are rapidly becoming part of students' everyday problem-solving toolkit in programming courses~\cite{hanifi2023chatgpt}. While these tools promise to democratize access to support and feedback, their unregulated use in educational settings raises fundamental questions about learning, trust, and correctness, especially in domains that demand rigorous reasoning, such as formal verification.

Formal verification tools like Dafny require users to write correct code, to formulate precise specifications, and to construct proofs. These tasks demand a solid  understanding of logical invariants (properties that must always hold true) and tool behavior. This creates a steep learning curve for students~\cite{noble2022more}, particularly those new to formal methods. %
It is in these cognitively demanding scenarios that LLMs may be most helpful (or not) as a support mechanism.

Prior work has explored how LLMs support or hinder novice programmers in writing code~\cite{qureshi2023exploring} or debugging simple scripts~\cite{groothuijsen2024ai}. However, much less is known about how students engage with LLMs during semantically complex tasks like deductive verification, 
where correctness requires not only syntactic accuracy but also logical soundness.
To address this gap, we conducted a controlled study with master's-level students enrolled in a formal methods course using Dafny. Students completed verification exercises with and without access to a custom ChatGPT interface. Through a survey, logs, and artifact analysis, we answer the following questions: %

\begin{description}

    \item[RQ1.] \textbf{[Performance]} Does ChatGPT improve students' performance in verification problems?

    \item[RQ2.] \textbf{[LLM Interaction]}\\ \change{}{RQ2.1:} How do students interact with ChatGPT when solving deductive verification problems?\\ 
    \change{}{RQ2.2:}  
    What strategies do more successful students employ when solving Dafny problems?

    \item[RQ3.] \textbf{[Trust]} How do students perceive and trust the responses provided by the LLM?
\end{description}

To our knowledge, 
this is the first 
empirical study of how students use LLMs to solve deductive verification problems in Dafny. Our main contributions are:
\begin{itemize}
    \item we show that LLMs significantly improve student performance, particularly on implementation tasks with Dafny;
    \item a detailed qualitative analysis of prompting strategies that distinguish more successful students;
    \item three actionable recommendations for using LLMs in the classroom.
\end{itemize}

\section{Background and Related Work}

\paragraph{\bf Formal Verification and Dafny.}

Dafny~\cite{leino2010dafny} is a verification-aware programming language %
designed to integrate formal verification into the software development workflow seamlessly. It enables developers to specify program behavior using constructs such as \texttt{requires}, \texttt{ensures}, and \texttt{assert}, which are then automatically verified by the Dafny tool. This process helps ensure that the code meets its intended specifications, reducing the risk of defects that might otherwise go undetected through conventional testing.
In industry, Dafny has been adopted across multiple domains, such as AWS~\cite{rungta2022billion} and software verification competitions~\cite{farrell2021using}.
Dafny is also an effective educational tool for introduction to formal methods, having been used, for example, for programming-intensive learning with immediate formative feedback via automated assessments~\cite{noble2022more}.

\paragraph{\bf LLMs and Dafny.}
Recent work has begun to integrate the capabilities of LLMs with formal verification workflows. One approach is to generate full code and specification from a function description in natural language. 
Clover~\cite{sun2024clover} and Mirchev et al.~\cite{mirchev2024assured} used LLMs like ChatGPT to generate code and specifications from natural language and to detect inconsistencies among code, specs, and documentation.
Similarly, Misu et al.~\cite{misu2024towards} explored different prompting techniques %
to improve the base performance of pretrained models. \change{}{Wu et al.~\cite{wu2025Dafny-APR-LLM} use LLMs for program repair in Dafny.} %

While some works focus on code generation, others focus on specification tasks. Silva et al.~\cite{silva2024leveraging} explored ChatGPT's ability to assist with lemma discovery and proof \change{outline}{} generation in Dafny \change{. They}{and,}  found early difficulties in producing syntactically correct code. Other tools aim to automate the inference of helper annotations, such as missing assertions or loop invariants, crucial for verification~\cite{mugnier2025laurel,poesia2024dafny,pascoal2025automatic}. %
For example, %
Pascoal et al.~\cite{pascoal2025automatic} focused on loop invariant synthesis, with the tool generating correct invariants on the first try in 92\% of cases and within five attempts in 95\%, leaving only the most complex examples unsolved.

\paragraph{\bf LLMs in CS Education.}

The use of LLMs in CS education has mixed outcomes. Groothuijsen et al.~\cite{groothuijsen2024ai} examined a scientific computing (C/Python) course for engineering students. They found that students used ChatGPT extensively for tasks like debugging and optimization, \change{but instructors noted reduced learning and pair programming}{but instructors observed declines in code quality and collaborative programming as students relied heavily on ChatGPT}. %
Sun et al.~\cite{sun2024would} study had similar findings, with no significant performance difference between ChatGPT-assisted and self-directed groups. %
Contrasting, Qureshi~\cite{qureshi2023exploring} ran a controlled lab in a data structures course: one team used only textbooks, and another could use ChatGPT. The ChatGPT group solved more test cases (higher scores) but also produced buggy or inconsistent code. The study concluded that ChatGPT provides a performance advantage on short programming challenges, but brings trade-offs (inaccuracies) that instructors must consider. 
Survey studies of software engineering students reveal high adoption but mixed trust. Hanifi et al.~\cite{hanifi2023chatgpt} surveyed SE undergraduates and found high usage and some self-taught prompt engineering, yet low overall trust, with 90\% of students reporting LLM hallucinations.
Prather et al.\cite{prather2024widening,prather2023s} found that stronger students used AI to accelerate progress, while weaker ones became over-reliant, often mistaking output for understanding.
Xue et al.~\cite{xue2024does} found no improvement in task outcomes and noted that ChatGPT use often displaced traditional learning without a clear benefit.
Overall, LLMs can aid students, but over-reliance may hinder learning and widen existing skill gaps.

\paragraph{\bf LLMs in Formal Methods Education.}

Formal methods are an important part of computer science curricula \cite{broy2024does}. To our knowledge, only two studies have evaluated LLM in formal methods education. Capozucca et al.~\cite{capozucca2025ai} examined ChatGPT's impact on students learning the B-method. A pretest-posttest experiment showed that access to the AI assistant did not improve the correctness of B-specifications. In fact, students who relied less on ChatGPT performed better, with lower trust in the assistant correlating with higher-quality results. Although ChatGPT helped surface key concepts, like state variables, it often failed to produce correct formal expressions. %
Similarly, Prasad et al.~\cite{prasad2023generating} integrated ChatGPT into a formal-methods course using Alloy/Forge. %
Over the semester, 64 students submitted only 293 prompts, mostly concentrated during an initial lab session. Of these, 81\% were course-related, and ChatGPT's responses were deemed relevant 86\% of the time. Despite this, survey data revealed a general reluctance to rely on the tool: 57\% of students feared that over-reliance on ChatGPT might hinder their learning, and 27\% expressed concerns about potential violations of course rules. Many students preferred solving problems independently, believing it would lead to better learning outcomes. However, Prasad et al.~\cite{prasad2023generating} study did not look at prompting proficiency or students' performance with the LLM. 
Our study is the first to explore in an educational context the use of LLMs \change{in the context of}{for} deductive verification, focused on Dafny.%

\section{Method}

\subsection{Study Design}

\begin{figure}[htbp]
  \centering

    \includegraphics[width=\linewidth, trim=0 11 0 10, clip]{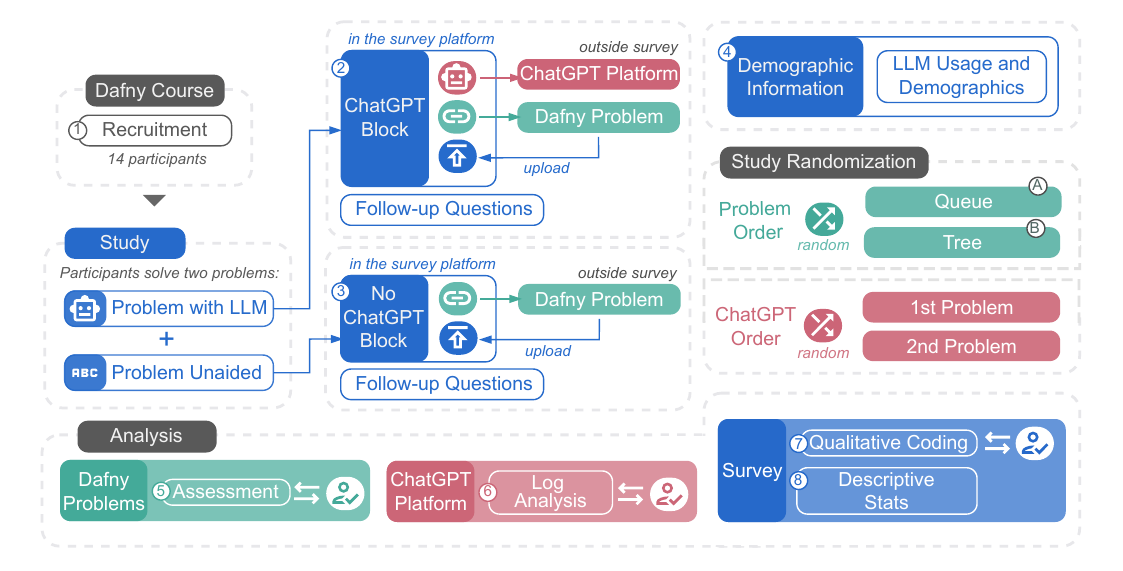}

  \caption{Overview of the study's methodology}
  \label{fig:method}
\end{figure}

Our \change{}{mixed-methods} study consisted of three main components: a Qualtrics-based survey, two domain-specific Dafny problems, and a custom ChatGPT interface designed for log collection. Figure~\ref{fig:method} illustrates the overall structure of the study. %

\paragraph{Survey.}  
We deployed the survey using Qualtrics. The survey served as the study's entry point and central coordination tool. Participants used the survey to provide informed consent, receive instructions, and access the exercises. After completing each problem, participants returned to the survey to upload their work and respond to follow-up questions. We designed the survey to preserve participant anonymity throughout. %

\paragraph{Problems.}  
Each participant completed two Dafny problems: \emph{Queue} and \emph{Tree}. We developed these problems iteratively over five pilot runs. Each problem consisted of \change{two subproblem}{two subproblems} and reflected realistic and moderately challenging tasks. The problems were designed to be solved in 30 minutes and to be LLM-resistant~\cite{mcdanel2025designing}. %
The two problems were slightly different from each other to mitigate learning effects, and their order was randomized per participant. Participants had a maximum of 30 minutes per problem and were instructed to download, complete, and re-upload the problem files using the lab computers provided during the session.

Both problems are based on a class that represents a data structure, namely a \textit{queue} implemented using a circular buffer, and a binary search \textit{tree}. Each provides the base code for the class, including the concrete and ghost fields, a constructor, and the \texttt{Valid} predicate. %

Each problem had a subproblem whose goal was to implement a method given its formal pre- and post-conditions, and another which gave them the natural language goal of the method and expected them to specify and implement it. The concrete subproblems are as follows, noting that the methods \texttt{Dequeue} and \texttt{Cut} already have their pre- and post-conditions defined for the participants:
\begin{itemize}
    \item \texttt{CircularBuffer<T>}'s \texttt{Dequeue}: a simple method \change{,}{}that should pop the first value from the queue and return it, naming it \texttt{elem}; %
    \item \texttt{CircularBuffer<T>}'s \texttt{Queue}: a method whose goal is to push the received value to the end of the queue;
    \item \texttt{BSTNode}'s \texttt{Cut}: a method that receives a maximum height and removes all the nodes whose height exceeds that value; %
    \item \texttt{BSTNode}'s \texttt{AddAll}: a method whose goal was to add all elements of a sequence to the binary search tree.
\end{itemize}

To ensure comparability between the two problems, we conducted a post-hoc statistical analysis of participants' performance in the \emph{Queue} and \emph{Tree} problems. 
We used a paired-sample t-test to compare the grades for the \emph{Queue} and \emph{Tree} problems across all participants, regardless of LLM condition, and found no statistically significant difference in difficulty between the two problems ($t = -0.653$, $p = 0.525$, \change{}{$d = 0.28$, }95\% CI: $[-7.58, 4.06]$). This result supports the construct validity of our study.

\paragraph{ChatGPT Platform.}  
We developed a custom web-based interface to provide controlled access to ChatGPT. This interface replicated the standard ChatGPT experience but recorded all user prompts and model responses. Each participant was randomly assigned to use ChatGPT for either the first or second problem. Those assigned to the ChatGPT condition received an additional link directing them to our custom platform and were asked to submit a key in the survey to link their log data with their responses.

The model used was \texttt{gpt-4\_1-2025-04-14} with default parameters. We chose to develop our own interface for two main reasons: (1) it ensured that all participants interacted with the same model under identical interface conditions, and (2) it simplified prompt logging and helped preserve participant anonymity. %
After logging in, each student received a random ID to enter in the survey. This ensured anonymity while allowing us to link ChatGPT logs to survey responses.

\paragraph{Experimental Workflow.}  
 All participants solved both problems but had access to ChatGPT in only one of them. We implemented a bucketed randomization procedure to assign participants to conditions. Rather than using full random assignment, we balanced the sample so that half of the participants performed each problem with and without ChatGPT. 
 We used the same method to randomize the order in which ChatGPT appeared.
 Both problem survey blocks (\blueCircled{2} and \blueCircled{3} in~\Cref{fig:method}) included 
 (1) instructions and a timer, 
 (2) a link to download the problem, 
 (3) a file upload field for submitting their work, and 
 (4) follow-up self-assessment questions. 
 The ChatGPT block (\blueCircled{2} in~\Cref{fig:method})
also included access to the custom ChatGPT interface and extra questions targeting the perceived usefulness of ChatGPT for different aspects of the task. A final section collected general demographic information and participants' broader experiences with LLMs (\blueCircled{4} in~\Cref{fig:method}).

\paragraph{Experimental Conditions.}  
Participants worked in computer labs without external internet access, using only a browser for the survey, our wrapper for ChatGPT,\change{}{ and the official Dafny documentation}. Each participant used Visual Studio Code equipped with the official Dafny extension and Dafny v4.10.0 installed locally. %

\subsubsection{Recruitment}
We recruited 14 participants (\greyCircled{1} in~\Cref{fig:method}) from the 2024/2025 cohort of the master's-level elective course \emph{Formal Methods for Critical Systems} at the Faculty of Engineering, University of Porto, Portugal\change{}{, taught by the last author. We recruited participants at the end of the semester after lectures had ended.} %
This course is part of a \change{two-year}{}Master's\change{programme}{} in Informatics and Computing Engineering and is offered during the second term of the first year of the degree.

\begin{table}

 \label{tab:demographics}
 \small
 \centering
 \begin{tabular}{llr@{\hskip 6pt}r@{\hskip 16pt}llr@{\hskip 6pt}r}
   \toprule
  {} & ~ & \textit{n} & \% & {} & ~ & \textit{n} & \% \\
  \toprule
  \multicolumn{4}{l}{\textit{Gender}} & \multicolumn{4}{l}{\textit{Education}} \\
  Male & & 12 & 86\% & Bachelor's degree & & 12 & 86\% \\
  Female & & 2 & 14\% & Graduate degree & & 1 & 7\% \\
  & & & & Some college, no degree & & 1 & 7\% \\
  \midrule
    \toprule
  {} & ~ & \textit{n} & \% & {} & ~ & \textit{n} & \% \\
  \toprule
  \multicolumn{4}{l}{\textit{Dafny LLM Usage}} & \multicolumn{4}{l}{\textit{AI Tools Used}} \\
  Every time & & 2 & 14\% & ChatGPT & & 10 & 71\% \\
  Most times & & 5 & 36\% & GitHub Copilot & & 5 & 43\% \\
  Sometimes & & 5 & 36\% & Gemini & & 3 & 21\% \\
  Rarely & & 2 & 14\% & Other & & 4 & 29\% \\
  \bottomrule
\end{tabular}

  \caption{Participants Demographics}
\end{table}

The course spans 14 weeks, with a weekly three-hour session, where about half of the time is used for theory and resolution of practical challenges in a tutorial style, and the other half is used for solving practical exercises independently. The first half of the course is dedicated to software modelling using Alloy, while the second half focuses on deductive verification using Dafny. 
Our study focuses on the Dafny component of the course. 

Participation was entirely voluntary, anonymous, and uncompensated. It was explicit that involvement would not influence academic performance or standing. The cohort comprised 16 students \,---\, the largest enrollment since the course was introduced in 2021/2022. %
Of these, 14 students voluntarily agreed to participate in the study, resulting in an 87.5\% participation rate (\greyCircled{1} in~\Cref{fig:method}).

The majority of participants identified as male (n = 12), and a smaller part as female (n = 2). In our sample, most participants had a Bachelor's or a Graduate degree (n = 13). Two participants reported using LLMs every time they worked on Dafny exercises, ten said they used it most times or sometimes, and two used it only rarely.
When asked about the AI tools they use (\Cref{tab:demographics}), most participants (n = 10) reported using ChatGPT and GitHub Copilot (n = 5).

\paragraph{Pilots.} We conducted five pilot sessions to refine the materials and procedure before deploying the study. Three of the pilot participants were former students who had previously completed the same Dafny course, and the remaining two had relevant background knowledge of Dafny. We conducted pilots until we reached saturation, that is, until no new suggestions or usability issues emerged. We used pilot data solely for internal feedback and did not include it in the final analysis.

\subsection{Analysis}\label{sec:analysis}

We did qualitative analysis using a double-blind, iterative coding process for the correctness assessment procedure and for both the prompt analysis and survey codebooks. Initially, one researcher conducted a first round of open coding and developed a preliminary version of a codebook. This preliminary codebook was then handed over to a second researcher, who independently coded the data and revised the codebook. This iterative process continued until the codebook stabilized and the coders achieved consensus.
The full annotated codebooks are included in the supplementary materials\footnote{\href{https://drive.google.com/drive/folders/1iAIkLP8hOPxZjAVss-lQghoaKiAEARU0}{Link to Supplementary Material}}.

\subsubsection{Correctness assessment procedure}

All participants’ solutions to the subproblems (four in total, two for each problem) were evaluated on a 0–20 point scale (rounded to two decimal places) according to the following criteria:

\begin{itemize}
    \item First, we divided each subproblem into tasks, which described the steps needed and expected of their solutions. Each subproblem has its own set of tasks, not necessarily the same number of them;
    \item We gave weights for these tasks according to their effort and difficulty. We also added the possibility of a penalty of, at most, one point, to be used in case the participant incorrectly added extra steps that made the code incorrect (e.g., an unnecessary additional contract);
    \item For each participant, we extracted each line of code and catalogued it according to the task they were trying to achieve;
    \item For each of those, we gave a grade from 0 to 20, according to how well the task was performed.%
\end{itemize}

In Dafny, the \emph{specification}\,---\,which includes the contracts, invariants, and ghost variable updates \change{}{(variables used only for verification purposes)}\,---\,defines precisely the expected behaviour of a program, and the \emph{implementation} provides the functionality for the program. So, we defined each task to be exclusively part of the specification or the implementation, also providing a way to calculate their grades for those specific parts of the methods.

Each participant has a grade for each problem, calculated with the unweighted average of their composing subproblem. 

\subsubsection{Prompt Analysis}

Our analysis focused on categorizing user prompts and ChatGPT responses into three categories:
\begin{enumerate}[label=(\arabic*), ref=\arabic*]
    \item \label{cat:prompt} Prompt characteristics (e.g., type and content): Captures prompt characteristics (e.g., structure and content), revealing how users initiate interactions.
    \item \label{cat:retry} Retry strategies (i.e., repair techniques): Covers retry strategies, showing how users revise or reframe prompts after unsatisfactory responses.
    \item \label{cat:interaction} Evaluation of the overall user interaction: Summarizes overall interaction patterns, including LLM reliance, emotional cues, and how closely users followed the tool's output.
\end{enumerate}

\change{We used an emergent coding process to analyze these responses as described in~\Cref{sec:analysis}.}{} %

\subsubsection{Survey Analysis}
We used quantitative and qualitative methods in our study. 

\paragraph{Statistical Analysis.}
We conducted descriptive and statistical analyses to evaluate the impact of ChatGPT assistance on student performance and self-reported confidence. Each participant completed two problems: one with ChatGPT assistance and one without. We analyzed the paired grades using a paired-sample t-test (after a Shapiro-Wilk's test to guarantee normality) to assess whether ChatGPT influenced academic performance. %

\paragraph{Qualitative Analysis.}
Our analysis focused on the open-ended responses collected through the survey, particularly in the section following the ChatGPT-assisted problem. The main open question was why participants trusted (or not) ChatGPT's responses for the problem. 
We developed the codebook with an emergent coding process. %

\paragraph{Ethical Considerations.}
We did not collect any personal or identifiable data, and all participants were over 18 years of age. Participants were clearly informed that they were not required to participate and could withdraw at any time without penalty. To ensure the activity was also beneficial to participants, we made the problems and their corresponding solutions available afterwards to all students enrolled in the course.
Prior to taking part in the survey or interview, all participants reviewed an informed consent form outlining the purpose of the study, the nature of their involvement, and their rights as participants. Only those who gave explicit informed consent were allowed to proceed. 
According to our institution's ethical guidelines, research with these characteristics, voluntary participation, no collection of personal data, and non-invasive procedures, does not require prior approval from the Ethics Committee. Nonetheless, we adhered to the principles outlined in the European Commission's Ethics Self-Assessment guide to ensure compliance with best practices in ethical research involving human participants.

\section{RQ1 Performance}

\paragraph{LLM effect on performance.}
The average grades using and not using the LLM were $17.39$ and $9.36$, respectively. 
All participants achieved a passing grade when using the LLM, whereas only 5 of the 14 achieved a passing grade without it %
(\emph{P2}, \emph{P3}, \emph{P6}, \emph{P10}, and \emph{P11}). Also, only one participant (\emph{P2}) worsened their grade with the LLM. All of these suggest that LLMs help improve the participants' grades.
We used a paired-sample t-test to compare grades from the solutions with LLM help to those without, and it showed a statistically significant difference in performance ($p = 0.0002$). On average, participants scored 5.04 points higher (\change{}{$d = 1.55$,} $95\%~CI: [4.59, 11.47]$) when accessing ChatGPT.

\paragraph{LLM effect on specification and implementation.}

The first and second subproblems of each problem have the same goal: the first is to implement a method, given its specification, and the second is to specify and implement a method from its natural language description. 
Comparing participants’ scores for each subproblem, the aggregated averages with the LLM were 19.52 for the first subproblem and 15.26 for the second, compared to 9.90 and 8.82 without the LLM.

We applied a paired t-test to assess whether the observed difference in performance between specification and implementation tasks was statistically significant, specifically on the second subproblem (with LLM), which consisted of both types of task.  One outlier was removed to meet normality assumptions, which were supported by a Shapiro-Wilk test ($W = 0.895$, $p = 0.116$). The test revealed a statistically significant difference in performance ($p = 0.003$), with participants scoring on average 2.92 points lower on the specification task (\change{}{medium effect size $d = 0.54$,} $95\%~CI: [-4.64, -1.19]$).

\paragraph{LLM effect on confidence.}

We now look at a particular student case, \emph{P5}, who scored zero on the problem without LLM assistance (which was their first problem) and 19 points with the LLM. In the non-LLM problem, the participant did not write a single line of code. Instead, they left comments indicating some steps of the solution, and also a comment that they felt unable to solve it in the time available because they were not familiar with the grammar. This self-assessment is surprising, as the participant had just completed the Dafny component of the course, including its evaluation. Interestingly, even in the LLM-assisted problem, where \emph{P5} produced a solution that scored nearly full points, the participant reported low confidence in their solution. From the answers to the survey, it was clear that the participant was not confident at all with his Dafny skills. It is possible that the fact that their solution did not verify contributed to this continued lack of confidence, despite their strong objective performance.

This case also shows that, for students who are unable to start solving a task, like \emph{P5}, who wrote no code unaided,  LLMs show promise in supporting them through the initial phases of problem-solving. However, it also reinforces a key limitation: LLMs cannot replace the need for conceptual understanding. While they can produce code that is syntactically and semantically correct, or near correct, learners may still feel insecure if they do not grasp the underlying logic, particularly in a formal system like Dafny.%

\begin{tcolorbox}[myboxstyle]
The usage of LLMs helped improve the grades of the students. However, for some, its absence was a source of insecurity, which shows that they must be careful when using these tools.
Also, concerning the performance of ChatGPT for Dafny, our results suggest that it is better for implementation tasks than specification ones. The properties of code currently available online, which predominantly does not include contracts or specifications, can explain this difference.
\end{tcolorbox}

\section{RQ2 LLM Interaction} 
In total, the ChatGPT interface collected 206 messages: 103 student prompts and 103 ChatGPT responses. To explore how students interact with ChatGPT during deductive verification tasks, we first analyze the coded prompts. The supplementary material provides the full codebook, along with category counts and user distributions. \change{}{This section is divided into two sub-research questions}.

\subsection{\change{}{RQ2.1 How do students interact with ChatGPT when solving deductive verification problems? }}

\paragraph{Prompt characteristics.} 
Twelve students used instruction-based prompts (34 total), and eight question-based prompts (23 total), as students could use both in the same prompt. Most (12/14) included the full class definition in at least one prompt to provide necessary context. %

\paragraph{Retry Strategies.}
The most common repair strategy was \emph{adding error messages} to the prompt, which was used by twelve students in 43 prompts, followed by \emph{including problematic code lines} used by 12 students in 26 prompts, and finally by \emph{redirecting the LLM} used by seven students in eight prompts. Redirecting refers to suggesting a specific approach for the LLM to use in solving the problem.

\paragraph{Overall User Interaction.}
Six students relied solely on prompting without modifying the \change{LLM code}{LLM-generated code}. Four made minor edits to LLM \change{output}{generated code}, while another four were largely autonomous, using the LLM sparingly \change{and heavily modifying code}{and performing manual modifications}. %
Two students started with partial implementations. In terms of emotional engagement, three students showed signs of frustration (e.g., P2 prompting the LLM  with ``did not help!'' after two repeated wrong responses) during the interaction, while two expressed clear satisfaction (e.g., P6 prompting LLM after a correct response ``it was strengthening the invariants, thank you'').

\begin{tcolorbox}[myboxstyle]
Most students were able to prompt the LLM effectively by providing sufficient context, such as including the full class definition, which proved critical for solving deductive verification tasks. Error-driven prompt refinement was the most common repair strategy, particularly by adding error messages and problematic code lines to guide the LLM. While nearly all students managed to extract either correct functionality or specifications from ChatGPT, only a subset (7/14 for subproblem 1 and 2/14 for subproblem 2) succeeded in obtaining fully verified solutions without significant manual intervention. 
\end{tcolorbox}

\subsection{RQ2.\change{1}{2} What strategies do more successful students employ when solving 
Dafny problems?}
\begin{figure}[htbp]
  \centering
    \includegraphics[width=0.9\linewidth, trim=0 11 0 10, clip]{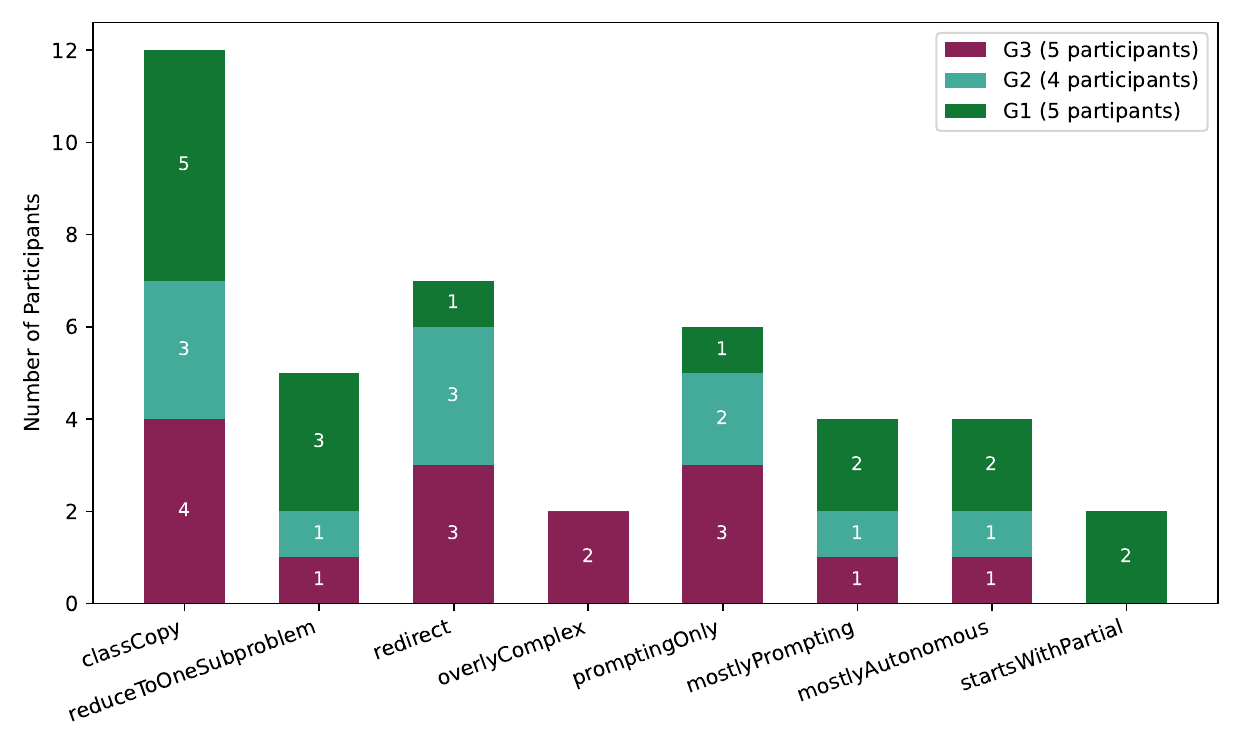}  
  \caption{Qualitative comparison of interaction patterns across participant groups (G1–G3), based on selected codes that reflect bigger differences between groups.}  %
  \label{fig:promptTypePerClass}
\end{figure}

To explore this, we divided participants into three groups based on their performance in the LLM-based exercise. \textbf{G1} includes the top five performers (scores 19 \change{through}{and} 20), \textbf{G2} the next four (scores 17 \change{through}{and} 18), and \textbf{G3} the remaining five participants (scores 10 through 16) that had the lowest scores. In~\Cref{fig:promptTypePerClass} we show the interaction metrics for each group. %
Due to space restrictions, only the codes (from the codebook) mentioned in this paper are detailed in \Cref{tab:prompt}. For information about all codes, please consult the supplementary materials\footnote{\href{https://drive.google.com/drive/folders/1iAIkLP8hOPxZjAVss-lQghoaKiAEARU0}{Link to Supplementary Material}}.

\begin{table}[ht]

\label{tab:prompt}
\begin{tabular}{@{}ll@{}}
\toprule
\textbf{Code} & \textbf{Description} \\ \midrule
\texttt{classCopy} & User copies the entire class in the prompt \\
\texttt{reduceToOneSubproblem} & Focuses on a single subproblem or method \\
\texttt{redirect} & Guides the LLM or teaches Dafny concepts \\
\texttt{overlyComplex} & Includes unnecessary complexity (lemmas, aux vars) \\
\texttt{promptingOnly} & User only interacted with LLM until solution \\
\texttt{mostlyPrompting} & Mostly used LLM, with minimal manual edits \\
\texttt{mostlyAutonomous} & Mostly implemented code independently \\
\texttt{startsWithPartial} & Started with partial user implementation \\
\bottomrule
\end{tabular}
\centering
\caption{Codes used in the qualitative comparison of interaction patterns}
\end{table}

\paragraph{Prompt characteristics.} 
Analysis of \textbf{classCopy} shows that all five participants in G1 included class definitions in their prompts, compared to three out of four in G2 and four out of five in G3. Providing this context was often essential for the LLM to generate a correct solution. For instance, \emph{P14} from G3 consistently failed to include sufficient information about the class and its methods, forcing the LLM to make assumptions about the surrounding code and ultimately leading to incorrect outputs. 

\paragraph{Retry Strategies.} 
A notable difference between groups emerged in the use of \textbf{redirect} strategies. Only one participant in G1 attempted to redirect the LLM, compared to three participants in both G2 and G3. These redirections often led the LLM to engage in unnecessarily complex reasoning, such as introducing superfluous lemmas, without actually addressing the root problem. This was particularly evident in the case of \emph{P9}, who persistently asked the LLM to use lemmas after it became stuck. Instead of resolving the issue, these redirections led to increasingly convoluted responses, ultimately keeping the LLM stuck. Consequently, two G3 participants were labeled as producing \textbf{overlyComplex} outputs, a pattern not seen in G1 or G2.

In contrast to the unproductive redirections seen above, \emph{P12} from G2 demonstrated a successful use of \textbf{redirect}. When the LLM became stuck while handling a ghost variable separation issue, P12 did not ask it to fix the code directly. Instead, they redirected the model by asking which variable could replace the problematic one. The LLM correctly suggested substituting `capacity' with `buffer.Length'. Though the model was unable to make the fix itself, the participant manually applied the suggestion and arrived at a correct solution. 

Interestingly, G1 participants more frequently employed the \textbf{reduceToOneSubproblem} strategy, which appeared in three cases, compared to only once in each of the other groups. This approach allowed the LLM to focus solely on a single subproblem, simplifying its overall task.
 
\paragraph{Overall User Interaction.}

User initiative varied noticeably across groups. In G1, only one participant relied solely on prompting (\textbf{promptingOnly}). Remarkably, this participant (\emph{P8}) solved both subproblems of the \emph{Tree} problem with a single prompt that included the entire Dafny file followed by the instruction \emph{``solve this''}. This level of success in a single attempt was not observed in any other group. A similar case occurred with \emph{P6}, also from G1, who used a slightly different prompt for the same problem. However, %
\emph{P6} did not recognize the correctness of the LLM's response and instead began directing it to revise the code. These examples illustrate the non-deterministic behavior of LLMs: identical or similar prompts can yield different outcomes, and even when a correct solution is generated, users may need sufficient domain knowledge to recognize and apply it.

In contrast, two participants in G2 and three in G3 also relied exclusively on prompting, but without the same level of success. G1 stood out for having more participants classified as \textbf{mostlyAutonomous}. It was also the only group in which two participants began with a partial implementation (\textbf{startWithPartial}), suggesting a proactive approach to structuring their interaction with the LLM.

\begin{tcolorbox}[myboxstyle]
Better outcomes were not solely determined by prompt quantity or persistence, but by strategic interaction choices. High-performing students were more likely to: (i) include class context early, (ii) reduce the prompt to ask only a single subproblem, (iii) avoid overcomplicating the LLM's task through redirection, and (iv) contribute with their own partial implementations. More successful students were also more autonomous, often making substantial modifications to the code provided by the LLM rather than relying on it verbatim. 
\end{tcolorbox}

\section{RQ3 Trust}

\begin{figure}[htbp]
  \centering
    \includegraphics[width=0.8\linewidth, trim=0 7 0 10, clip]{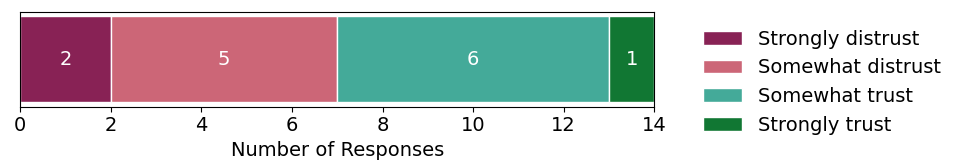}
  \caption{Participants' responses to whether they trusted ChatGPT during the exercise with a 5-point Likert scale} 
  \label{fig:trust}
\end{figure}

We asked participants if they trusted %
ChatGPT's responses during the Dafny exercise and why. Overall, participants expressed mixed attitudes. Of the 14 students, seven trusted the LLM and seven distrusted it (see~\Cref{fig:trust}).

\paragraph{Reasons for Trust.} The seven students who reported trust in ChatGPT cited several reasons. The most common was that it produced accurate code, mentioned by three participants. For example, P3,  ``Strongly'' trusted ChatGPT and mentioned \textit{``I resorted to ChatGPT and it solved my mistakes immediately.''}. Two students also mentioned that they trusted the LLM because they could independently verify ChatGPT's suggestions with Dafny built-in verification, as P4 described ``\textit{Because with Dafny it's easy to assert if its suggestion is correct or not}.''. The two remaining codes were only mentioned by one participant each. P9 mentioned that they trusted ChatGPT because it helped them ``understand key concepts'' and P12 mentioned ChatGPT was better than nothing, saying ``\textit{I don't feel that comfortable in Dafny, so in my case (...) I have to trust.}''

\paragraph{Reasons for Distrust.} Seven students distrusted ChatGPT, and of these, four distrusted it because of syntax errors, as it often produced invalid Dafny code that required manual correction (``\textit{responses (...) contain one or two little errors, they are sometimes easy to find (...) but [I] still need to be careful.}'', P8). Two participants distrusted the LLM because it was overconfident when wrong. One participant mentioned directly hallucinated features (``\textit{[it] kept insisting on wrong suggestions, sometimes suggested things that do not exist}'', P2). For one participant, distrust came from prior experiences with ChatGPT rather than the current study, and another said that ChatGPT is not useful for verification. %

\paragraph{What Influences Trust?}
Students' levels of trust may not be fully aligned with their performance. One example is P5, who scored zero without ChatGPT but had a near-perfect solution with it. Despite this improvement, the participant still distrusted the LLM's output. While ChatGPT can help students complete tasks more successfully, it does not automatically increase their confidence because if it produces correct code that students do not fully understand, they may remain skeptical, even in the face of strong objective performance.
On the other hand, some students appeared to place significant trust in ChatGPT %
without much critical evaluation. For instance, participant P3 noted that they were tired, turned to ChatGPT, and it produced code quickly, so they ``strongly trusted'' it.

\begin{tcolorbox}[myboxstyle ]
Students' trust in ChatGPT appears to be influenced by several factors: the technical correctness of responses, the presence of verification feedback (via Dafny), students' confidence in formal methods, and prior experience with AI tools.
\end{tcolorbox}

\section{Discussion}

In this section, we reflect on the broader pedagogical implications of our findings.

\subsection{Implications for CS Pedagogy }

\paragraph{\bf Success Does Not Equal Understanding.}
Our results suggest that good scores with LLM do not necessarily reflect deeper understanding or trust. We saw cases of participants who improved their performance with the LLM and still distrusted it. So, while LLMs can enable students to progress where they might otherwise stall, it does not automatically help learning or understanding.
This disconnect raises important questions about dependency, as students may rely on LLMs as a substitute for their reasoning, consistent with prior work~\cite{jovst2024impact,prather2024widening}. Our survey data offers preliminary support for this concern. Four of the five students who scored above 10 (out of 20) on the unaided %
problem reported using LLMs ``sometimes'' or ``rarely'' outside the study. In contrast, the one student in this group who used LLMs ``most of the time'' performed the worst. %

For \textbf{educators}, these findings present both a challenge and an opportunity. If LLMs are allowed in the classroom, instructors should monitor how students use them. Some students may achieve better scores while blindly accepting the LLM’s output, which can mask deeper learning issues. Identifying these patterns can serve as a diagnostic tool. Since all students in our study reported using LLMs outside the experiment, classroom tools designed to track LLM interactions (like the one we used) can help detect usage patterns and support a more targeted intervention.

\paragraph{\bf Prompt Design is Critical.}
Our findings suggest that the benefits of ChatGPT in Dafny exercises depend on how students use the LLM. High-performing students tended to write prompts with all relevant context and focused on solving one subproblem at a time. In contrast, lower-performing students often omitted crucial details and redirected the LLM to overly complex solutions. Students who experimented with and modified the LLM's output
made more progress than those who simply copied responses and relied on repeated prompting. These results align with prior research, where prompt quality determines LLM support's effectiveness in formal methods education~\cite{capozucca2025ai}. All participants reported using LLMs outside the classroom when working on Dafny exercises, so ineffective prompting practices may not be limited to the experimental setting and may be affecting students' independent learning as well.

For \textbf{educators}, this highlights the need to explicitly teach productive prompting strategies rather than simply allowing or discouraging LLM use. Structured prompting templates, or examples, could scaffold the learning process and help students internalize good practices over time.
The prompting strategies that proved effective in our study may be transferable to other programming tasks. Our results suggest that educators should encourage students to think critically before using LLMs, avoid overly complex solutions, and treat the LLM as a support tool. Breaking down larger problems into smaller steps also seems an effective strategy when working with LLMs.
Future work should explore whether teaching prompt design directly leads to improved learning outcomes and whether these benefits generalize across domains.

\paragraph{\bf Designing LLM-Resistant Challenges is not Trivial.}
One of the main challenges we encountered %
was developing problems that LLMs could not solve instantaneously. We wanted to write complex but solvable exercises and found that ChatGPT could immediately solve many complex exercises (e.g., from DafnyBench~\cite{loughridge2024dafnybench}). 
We designed our challenges from scratch, independent of existing online problems. We iteratively tuned problem complexity to balance difficulty and solution time with five pilot rounds. We found that LLM-resistant problems are challenging to design, even taking into consideration best practices~\cite {mcdanel2025designing}. %

For \textbf{educators}, our insights show that they must consider what students \textit{and} LLMs can do. Some design decisions in our problems contributed to increased LLM resistance (e.g., %
constrained the solution to our defined class invariants, predefined contracts, and selected ghost variables\,---\,which are not usable in the implementation). These elements reduced the degree of freedom in the solution space, a known strategy for LLM resistance~\cite{mcdanel2025designing}.

Our results revealed that ChatGPT struggled more with specification tasks than with implementation. Educators can design exercises focused on specifications (e.g., writing contracts) as these may be less vulnerable to LLM assistance. This insight also has broader implications beyond education. In \textbf{industrial} contexts, they motivate a shift towards specification quality and completeness. If a company uses an LLM assistant for verification, they should focus the human input on the specification and design of correctness properties rather than the implementation itself.

\subsection{Limitations}
As with all empirical studies, there are some limitations to our research. First, our sample size was relatively small (14 participants), reflecting the voluntary participation of students enrolled in a specialized formal methods course. We advise future work to look into replicating our protocol on a larger scale.
Some of our results can be affected by factors related to task sequence or prior familiarity with Dafny. %
To address this, we designed tasks of comparable difficulty, randomized the order of tasks, assigned ChatGPT access, and conducted multiple pilot sessions. %
Finally, our study focused on interactions with a single LLM, ChatGPT, accessed through a custom web interface. 
This allowed control over the user experience and interaction logs, but it may not fully represent students' real-world LLM use. 
Our findings may not generalize to other LLM platforms. %

\section{Conclusion}
\textit{Can large language models help students prove software correctness?} Our findings suggest that the answer is \textit{yes}, but with important caveats. Students performed significantly better with LLM support, especially in implementation tasks. However, LLM's effectiveness depended on students' interaction strategies. Success requires good prompt design and full program context. LLMs can be valuable assistants for proving software correctness, not as solvers but as collaborators whose utility depends on how students use them.

\begin{credits}
\subsubsection{\ackname} 
This work was financed by National Funds through the FCT - Fundação para a Ciência e a Tecnologia, I.P. (Portuguese Foundation for Science and Technology) within the project VeriFixer, with reference 2023.15557.PEX (DOI: 10.54499/2023.15557.PEX), and by funding from FCT under grants 2024.00375.BD, PRT/BD/153739/2021, and PRT/BD/155045/2024.

\end{credits}
\bibliographystyle{splncs04}
\bibliography{bib}

\end{document}